\documentclass{emulateapj}
\usepackage{apjfonts}
\def\gs{\mathrel{\raise0.35ex\hbox{$\scriptstyle >$}\kern-0.6em
\lower0.40ex\hbox{{$\scriptstyle \sim$}}}}
\def\ls{\mathrel{\raise0.35ex\hbox{$\scriptstyle <$}\kern-0.6em
\lower0.40ex\hbox{{$\scriptstyle \sim$}}}}

\newenvironment{inlinetable}{%
\def\@captype{table}%
\noindent\begin{minipage}{0.999\linewidth}\begin{center}\small}
{\end{center}\end{minipage}\smallskip}

\def\micron{$\mu$m\ }
\def\etal{et\ al.\ }

\lefthead{J. E. Geach \etal}
\righthead{SST observations of LABs}
\begin{document}

\title{{\it Spitzer} identifications and classifications of submillimeter galaxies in giant, high-redshift Lyman-$\alpha$ emission-line nebulae}

\author{J.~E.~Geach\altaffilmark{1}, Ian~ Smail\altaffilmark{1},
  S.~C.~Chapman\altaffilmark{2,3}, D.~M.~Alexander\altaffilmark{1},  A.~W.~Blain\altaffilmark{4},
  J.~P.~Stott\altaffilmark{1} and R.~J.~Ivison\altaffilmark{5}} 

\altaffiltext{1}{Institute for Computational Cosmology, Durham
  University, South Road, Durham DH1 3LE. U.K.\ j.e.geach@durham.ac.uk} 
\altaffiltext{2}{Institute of Astronomy, Madingley Road, Cambridge, CB3 0HA. U.K.} 
\altaffiltext{3}{CSA Fellow, U. of Victoria, Victoria BC, V8P 1A1. Canada} 
\altaffiltext{4}{California Institute of Technology, 1200 East
  California Boulevard, Pasadena, CA 91125. U.S.A.} 
\altaffiltext{5}{U.~K. Astronomy Technology Centre, Royal Observatory,
  Blackford Hill, Edinburgh, EH9 3H5. U.K.}

\begin{abstract}

Using {\it Spitzer Space Telescope} IRAC (3.6--8$\mu$m) and
MIPS (24$\mu$m) imaging, as well as {\it Hubble Space Telescope}
optical observations, we 
identify the IRAC counterparts of the luminous power sources residing within 
the two largest and brightest Lyman-$\alpha$ emitting nebulae (LABs) in the
SA\,22 protocluster at $z=3.09$ (LAB\,1 and LAB\,2). These sources are also both
submillimeter galaxies (SMGs). From their rest-frame optical/near-infrared
colors, we conclude that the SMG in LAB\,1 is
likely to be starburst dominated and heavily obscured
($A_V\sim3$). In contrast, LAB\,2 has excess rest-frame $\sim$2\micron emission (over that expected
from starlight) and hosts a hard X-ray source at the proposed location of the SMG, 
consistent with the presence of an active galactic nucleus (AGN). We conclude that LAB\,1 and LAB\,2 appear to 
have very different energy sources despite having similar
Lyman-$\alpha$ spatial extents and luminosities, 
although it remains unclear whether on-going star-formation or periodic
AGN heating is responsible for the extended Lyman-$\alpha$ emission. We find that the
mid-infrared properties of the SMGs lying in LAB\,1 and LAB\,2 are similar to
those of the wider
SMG population, and so it is possible that extended Lyman-$\alpha$ haloes
are a common feature of SMGs in general.
\end{abstract}

\section{Introduction}

Narrow-band surveys have revealed an intriguing new population: 
giant, radio-quiet Lyman-$\alpha$ emission-line nebulae (Steidel
\etal 2000), which have been termed `Lyman-$\alpha$ Blobs' (LABs). These
objects typically have linear extents of 10--100\,kpc, and
Lyman-$\alpha$ luminosities of 10$^{43-44}$\,erg\,s$^{-1}$, with the
highest concentration currently known in the SA\,22 proto-cluster at $z=3.09$,
with 35 confirmed objects (Matsuda et\ al.\ 2004). Similar objects have
been discovered
in other regions associated with rich, but primitive environments (Keel
\etal 1999; Francis \etal 2001; Palunas \etal 2004), and extended Lyman-$\alpha$
emission has been detected around certain
high-redshift radio galaxies (e.g.\ Reuland \etal  2003); however, the
narrow-band selected LABs are typically not associated with such radio-loud systems. 

Cooling-flows or shock heating by outflows originating from starburst
winds or AGN jets have been proposed as the most
plausible mechanisms for producing these extended Lyman-$\alpha$ haloes
(Taniguchi \& Shioya 2000; Ohyama \etal 2003), 
although other processes such as sub-sonic $pdV$ heating, and inverse Compton
scattering of Sunyaev-Zelodovich photons could also play a role (Scharf \etal
2003). 
The two largest known LABs (LAB\,1 and LAB\,2, with spatial extents of
1.1$\times10^4$\,kpc$^2$
and 0.8$\times10^4$\,kpc$^2$ respectively, both at $z=3.09$ in SA\,22; Matsuda \etal 2004) are both
associated with submillimeter galaxies (SMGs) (Chapman \etal 2001). 
Several other LABs in the SA\,22 structure at $z=3.09$ contain SMGs (Geach \etal 2005) 
-- the apparent association of these bolometrically luminous galaxies with
LABs provides support for the scenario where feedback mechanisms are powering
the Lyman-$\alpha$ emission, since the bolometric luminosities of the SMGs are
several orders of magnitude larger than the Lyman-$\alpha$ luminosities of the
haloes.
The action of superwind feedback has recently been given
credence by integral field data which reveals outflows and complex velocity
structures in both LAB\,1 and LAB\,2 (Bower \etal 2004; Wilman \etal
2005). However, the precise location and identification of the SMG
counterparts, and hence accurate constraints on the dominant power sources, are
unclear due to the poor spatial resolution of the submillimeter data
(Chapman \etal 2004).

In this Letter, we present {\it Spitzer Space Telescope (SST)} mid-infrared and {\it Hubble Space
Telescope (HST)} optical imaging
of the galaxies embedded within LAB\,1 and LAB\,2 to locate the bolometrically
luminous sources within them -- the SMGs. To identify these galaxies, we exploit
the fact that mid-infrared observations have been
shown to be effective at identifying SMGs, particularly at 8$\mu$m since these
galaxies have red infrared colors (Ashby \etal
2006). The mid-infrared photometry also constrains their rest-frame
optical--radio spectral energy distributions (SEDs) and
hence can shed light on what is powering these sources. 
Throughout we adopt $\Omega_m=0.3$,  $\Omega_\Lambda=0.7$ and
$H_0=75$\,km\,s$^{-1}$\,Mpc$^{-1}$. In this geometry, 1$''$ corresponds
to 7.1\,kpc at $z=3.09$. 

%
%
\begin{figure*}[t]
\centerline{
\includegraphics[width=3.in]{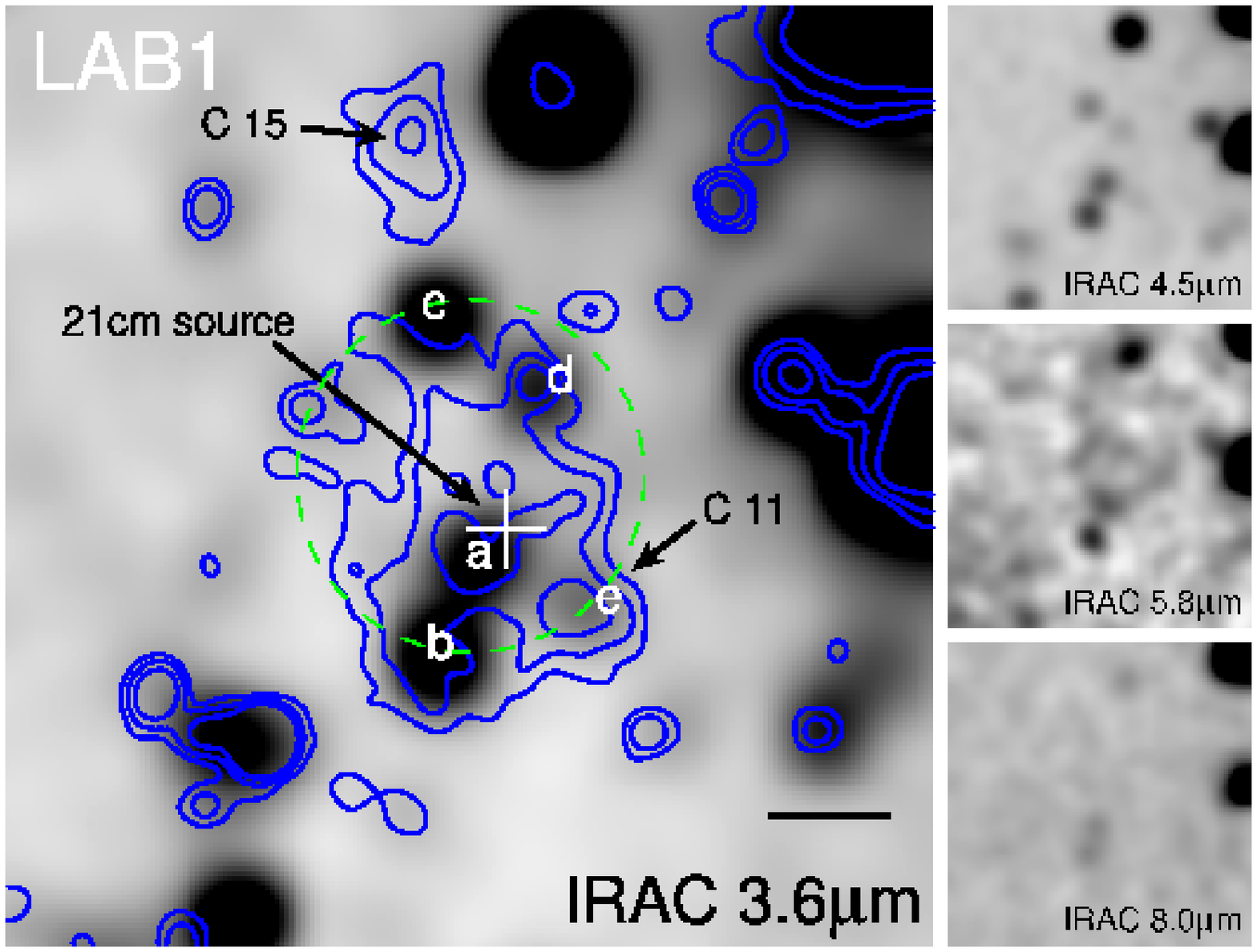}\hspace{3mm}\includegraphics[width=3.in]{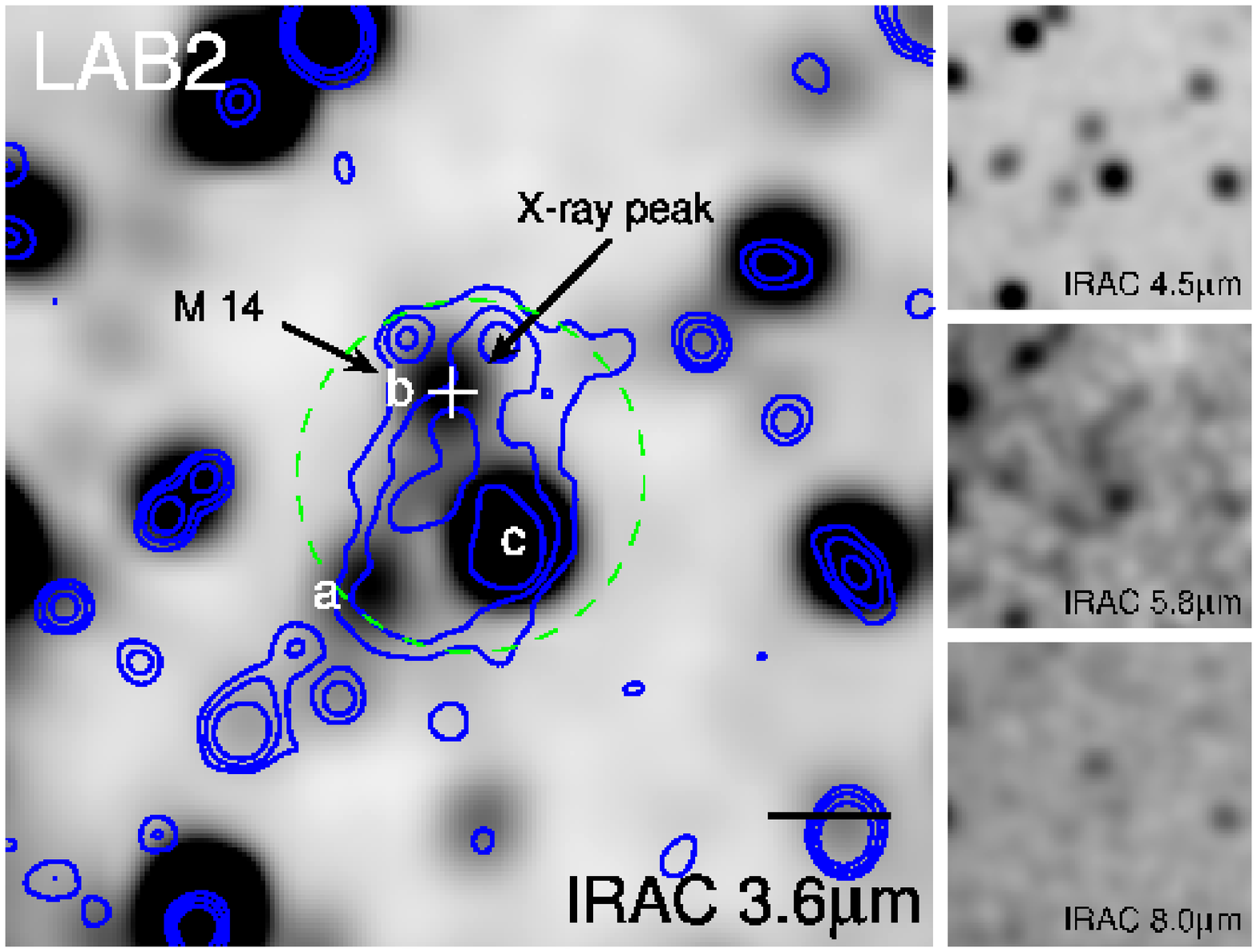}}
\vspace{1mm}
\centerline{
\includegraphics[width=3.in]{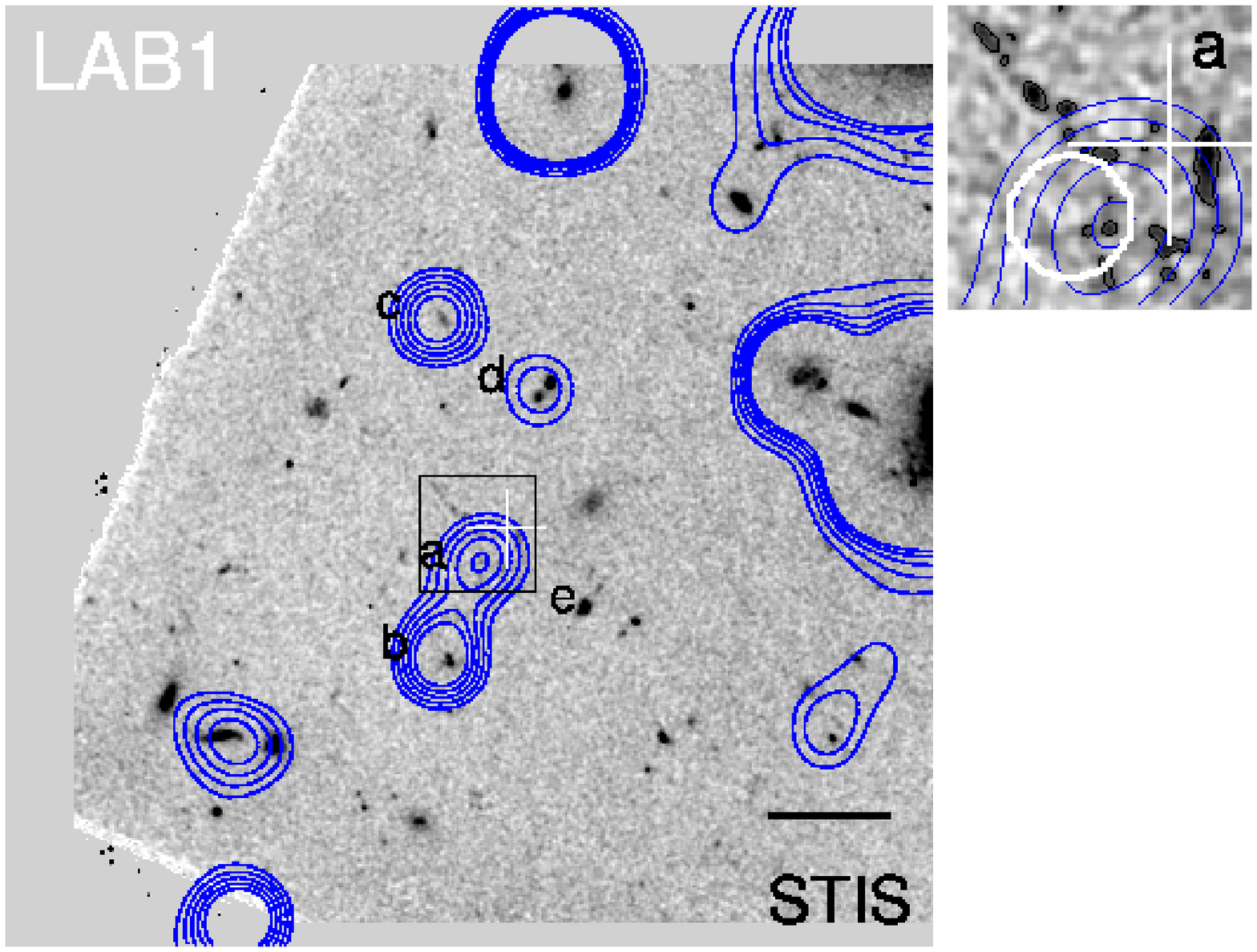}\hspace{3mm}\includegraphics[width=3.in]{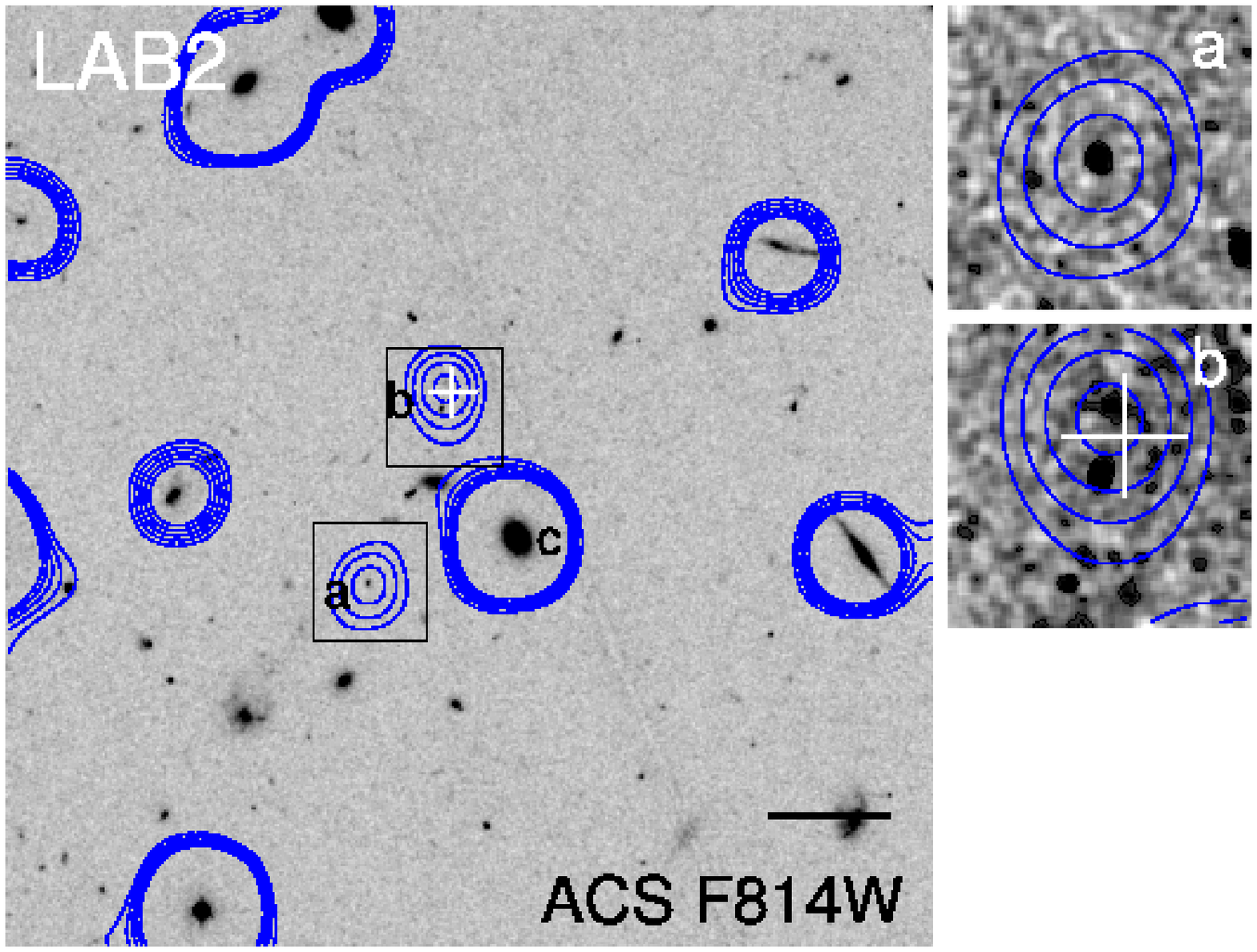}}
\caption{\small ({\it top}) Greyscale IRAC $40''\times40''$ images of LAB\,1
  ({\it left}) and LAB\,2 ({\it right}). The main panels show
  the IRAC 3.6$\mu$m image, and the insets show the 4.5$\mu$m 5.8$\mu$m
  and 8$\mu$m images, all which have been
  slightly smoothed with a Gaussian kernel. We overlay the narrow-band
  Lyman-$\alpha$ images from
  Matsuda \etal (2004) to show the extent of each halo. 
  IRAC detections within the Lyman-$\alpha$ haloes are labelled
  a, b, etc. In the field of LAB\,1 we indicate the location of
  the known Lyman-break galaxies (LBGs; C\,11 \& C\,15) and the position of the
  21\,cm radio source in the halo ({\it white cross}, Chapman \etal 2004). 
  LAB\,2 contains one known LBG
  (M\,14) and an SMG, as well as an X-ray source ({\it white cross}). 
  In both cases we show the 15$''$ SCUBA beam ({\it dashed circle}) to illustrate 
  the potential uncertainty in the position of the submm
  source. ({\it bottom}) High resolution {\it
    HST}-STIS and ACS imaging for LAB\,1 and LAB\,2 showing the optical
  counterparts, with contours showing the 3.6\micron emission. 
  The white 1$''$ circle in the STIS inset shows the location of the
  ERO detected in Steidel \etal (2000). 
  In all images, north is up, east is left and the horizontal bar 
  is a 5$''$ scale. We discuss the
  identifications further in \S3 and Table~1.} 
\end{figure*}

\section{Observations \& Reduction}
 
The {\it SST} observations used here are part of GTO project \#64, which
we retrieved from the {\it Spitzer} Science Center (SSC) archive. LAB\,1 and
LAB\,2 are covered by all four IRAC channels (3.6--8$\mu$m), and we 
use the Post Basic Calibrated
IRAC frames generated by the pipelines at the SSC. 
Both LABs were also observed by MIPS at 24$\mu$m as part of the same project,
and for these data we
perform post-processing on Basic Calibrated Data frames to remove
common MIPS artifacts and flatten small and large-scale
gradients using `master-flats' generated from the
data. For mosaicking we use the {\it SSC} {\sc mopex} package,
which makes use of the supplementary calibration files which are supplied
with the main science set. 
We use {\sc SExtractor} (v2.4.4, Bertin \& Arnouts 1996) to detect and
extract sources in the mid-infrared images. Fluxes are measured in apertures with 
diameters chosen to be the same as in the SWIRE ELAIS-N1 catalog\footnote{see Surace, 
J. A. et al.\ 2004, VizieR Online Data Catalog, 2255}: 3.8$''$ for IRAC and 12$''$ for
MIPS. 

In addition to the mid-infrared data, we analyse optical {\it HST} 
observations of the LABs, to reveal the rest-frame
ultra-violet (UV) morphologies of the galaxies in the LAB haloes. 
The STIS coverage of
LAB\,1 has been previously discussed by Chapman\ et\ al.\ (2004), while the 
Advanced Camera for Surveys (ACS) F814W ($I$-band) imaging
of LAB\,2 is presented here for the first time. The ACS data was reduced
using {\sc Multidrizzle}, and will be described in more detail
in J.~E.~Geach et al.\ (2006, in preparation).

%
%
\begin{figure*}[t]
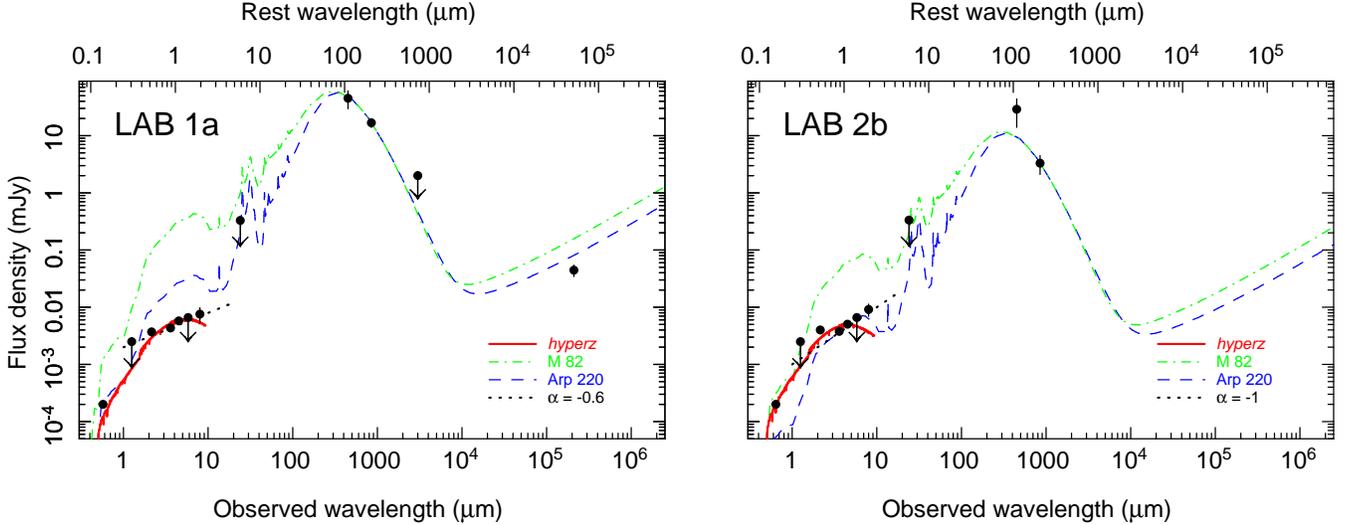

\centerline{
\includegraphics[width=3.5in]{f2a.ps}\includegraphics[width=3.5in]{f2b.ps}}  
\caption{\small Rest-frame ultraviolet to radio spectral energy distributions (SEDs) of
  the proposed power sources in LAB\,1 and LAB\,2. We overlay the SEDs of M\,82
  and Arp\,220 (Silva et al.\ 1998) normalized to the 850\micron point and 
  non-{\it Spitzer} data from the literature (upper limits at 3-$\sigma$). Both
  sources have broad-band SEDs which are similar to Arp\,220. Dotted lines
  show a simple
  power law, $f_\nu \propto
  \nu^\alpha$, fit to the 3.6--8\micron data. 
  We also show the best fit (stellar) SED from {\sc
  hyperz}, fitted to the optical/mid-infrared photometry (see \S4 for details).} 
\end{figure*}

\section{Identification of the SMGs}

As the precise location of the SMGs within the LAB haloes is uncertain, 
we begin our discussion by identifying the mid-infrared counterparts to the
SMGs (which we assume to be the dominant power-source of the LABs).
Starting with a catalogue of the IRAC-detections within each
Lyman-$\alpha$ halo\footnote{For mid-infrared detections within the Lyman-$\alpha$
haloes, we adopt the alphabetical suffix LAB\,1a, LAB\,1b, etc.,
although this does not imply that
these objects are physically associated with the LAB.} (Table~1), we aim to
eliminate sources not at $z=3.1$, and then identify the most likely
mid-infrared counterpart (if any) to the SMG in each case. Note that no 
sources are detected at 24$\mu$m within the extent of either LAB, 
and so we adopt 3-$\sigma$ upper
limits of $f_{24} < 330$\,$\mu$Jy.

Figure~1 shows {\it HST} and IRAC images of the LABs covering 3.6--8$\mu$m, and we
label each of our potential counterparts. Of the five IRAC components in
LAB\,1 (a--e), we can eliminate several of them as the SMG counterpart
immediately: LAB\,1c and d have mid-infrared colors $f_{3.6}/f_{4.5} >
1$, making them inconsistent with $z=3.1$ galaxies (Ashby \etal 2006),
while LAB\,1e is at the position of the Lyman-break galaxy (LBG) C\,11, and
is only weakly detected
at 3.6$\mu$m and undetected at $4.5$--$8\mu$m. Although this LBG is close
to LAB\,1, it appears to be kinematically distinct from the bulk of the halo
(Bower \etal 2004) and so we discount it -- leaving LAB\,1a and b as potential
counterparts. 

The positional uncertainty between LAB\,1a
and the 850\micron centroid ($\Delta\theta = 2.0\pm 1.9''$) is
consistent with it being the source of the submm emission. 
The remaining IRAC counterpart to the south-east of the SMG (LAB\,1b) is just
within the Lyman-$\alpha$ halo, and has a bright
counterpart in the STIS imaging. It is not clear from the mid-infrared
colors whether this galaxy is at the halo redshift, but given its large
separation from the nominal submillimeter position, it is unlikely that it is
associated with the power-source. However, if LAB\,1b is at $z=3.1$, it is 
possible that LAB\,1a and b are interacting given their projected separation of 
just $\sim$4$''$ ($\sim$30\,kpc at $z=3.1$).

Previous multi-wavelength observations of LAB\,1 (Chapman \etal 2004) 
associated the SMG with the brightest of a group of UV fragments near the 
geometrical center of the structure and close to the 21\,cm source position (Fig.~1). 
Component LAB\,1a is in the vicinity of this group of faint UV fragments, 
but the 3.6--8$\mu$m emission is coincident with a $K_S = 21.5$ 
Extremely Red Object (ERO) identified in Steidel \etal (2000), 
rather than the brightest UV component (`J1' in Chapman \etal 2004) which is 
closest to the radio peak (and a tentative CO(4$-$3) detection). 
Both LAB\,1a and the UV object J1 are plausible sources of the submm emission. 
However, given that SMGs are
expected to be massive, we view the 3.6--8$\mu$m
detection as a more reliable indicator of the SMG's counterpart than the weak UV
source. We therefore propose that LAB\,1a is the most likely IRAC counterpart to
the SMG in LAB\,1

There are three mid-infrared counterparts within the extent of  
LAB\,2 (Fig.~1, Table~1). 
We can eliminate LAB\,2c immediately as a potential counterpart given its blue
mid-infrared colors and large, bright optical counterpart in the {\it HST}
image -- it is a foreground object. The remaining two counterparts LAB\,2a and
LAB\,2b have
similar 3.6--5.8\micron characteristics, and similar compact {\it HST} morphologies
(Fig.~1). However, only LAB\,2b is detected at 8$\mu$m, which has been shown
to be a useful identifier of the mid-infrared counterparts of SMGs (Ashby \etal
2006). Moreover, LAB\,2 contains a
hard X-ray source in the halo (Basu-Zych \& Scharf 2004), with an
estimated luminosity $L_X\sim10^{44}$\,erg\,s$^{-1}$. 
The location of the X-ray source is
consistent with LAB\,2b ($\Delta\theta = 0.8\pm1.4''$), whereas
LAB\,2a is in the extreme south-east of the LAB, $\sim$8.5$''$ away from the
X-ray source. 
Given the association between SMGs and moderately bright X-ray sources
(Alexander et al. 2005a), the SMG in LAB\,2 is likely to lie close to
the position of the X-ray source. We therefore propose that LAB\,2b is the mid-infrared counterpart to the
SMG in LAB\,2\footnote{LAB\,2b is also at the location of the LBG
M\,14, although in this case the LBG appears kinematically associated with the
wider halo (Wilman et al.\ 2005). We note that in the submillimeter 
observation of this halo, LAB\,2b is actually situated near to the edge of the 
SCUBA beam (Fig.~1). The
correction to the 850$\mu$m flux to account for this offset would make the SMG
brighter, with $S_{\rm 850\mu m} \sim 5$\,mJy (Chapman \etal 2001).}.
The precise positions of the SMGs in LAB\,1 and LAB\,2 and the offsets to the
other mid-infrared components are given in Table~1.  

\section{Classification of the SMGs}

We can use the 3.6--8\micron photometry (along with multi-wavelength data for
these objects from the literature; Chapman \etal 2001,\ 2004; Steidel \etal
2000) 
to construct the optical--radio SEDs for the SMGs (Figure 2). 
We present new $J$ and $K$ band photometry from
the UKIDSS-Deep Extragalactic Survey (to be discussed in Stott et al.\ 2007 in
prep). We note that the poor
constraints on the $f_{\rm 24}/f_{\rm 850}$ ratio means that we cannot make
robust conclusions about the rest-frame mid-to-far infrared differences 
between LAB\,1a and LAB\,2b. Both have  $f_{\rm 24}/f_{\rm 850}$  upper limits
consistent with obscured ULIRGs at $z\sim3$ (Pope et al. 2006). Similarly the
limits of the observed $f_{5.8}/f_{850}$ ratios for LAB\,1a and LAB\,2b are consistent
with other SMGs at $z\sim3$, although the ratio for LAB\,1a is
at the low end of the distribution (Pope et al.\ 2006) indicating either a
relatively low stellar mass for this system, or more likely very high
obscuration. Thus, the SMGs
associated with LABs are similar to the wider SMG population.

We now examine the rest-frame near-infrared portion of the SEDs using the IRAC
data. Near-infrared colors have been shown to be a useful 
diagnostic for the power-source of local luminous infrared galaxies 
(e.g.\ Spinoglio \etal 1995).
For LAB\,1a we find rest-frame (Vega) colors 
$(J-H)=0.6\pm0.5$ and $(H-K) = 0.8\pm0.4$, and in LAB\,2b we find $(J-H) = 0.8\pm
0.6$ and $(H-K) = 1.1\pm 0.4$. LAB\,1a's colors are similar to that of
`normal' galaxies ($\left<J-H\right> \sim 0.8$,  $\left<H-K\right> \sim 0.3$;
Spinoglio \etal 1995), whereas LAB\,2b exhibits a redder rest-frame $(H-K)$
color, indicating an excess in the rest-frame 
$K$-band likely associated with a non-stellar contribution. 
We fit a simple power-law, $f_\nu \propto \nu^\alpha$, to the IRAC photometry for
LAB\,2b, obtaining $\alpha=-1.0\pm 0.2$, similar to that expected for AGN 
(Alonso-Herrero et al. 2006).

%
%
\begin{inlinetable}
\vspace{2mm}
{\sc Mid-IR properties of sources lying within LABs}
\vspace{2mm}
\begin{tabular}{lr@{.}lr@{.}lr@{.}lr@{.}lr@{.}lr@{.}l}
\hline
\hline
 & \multicolumn{4}{c}{(1)} & \multicolumn{8}{c}{(2)} \cr
 Component& \multicolumn{2}{c}{$\Delta\alpha$} & \multicolumn{2}{c}{$\Delta\delta$} &
 \multicolumn{2}{c}{3.6$\mu$m} & \multicolumn{2}{c}{4.5$\mu$m} &
 \multicolumn{2}{c}{5.8$\mu$m} &
 \multicolumn{2}{c}{8.0$\mu$m} \cr
 & \multicolumn{4}{c}{($''$)} &\multicolumn{8}{c}{($\mu$Jy)} \cr
\hline
LAB\,1{\bf a}\ldots\ldots\ldots& {\bf 0}&{\bf 0}    & {\bf 0}&{\bf 0}    & {\bf
4}&{\bf 3} & {\bf 5}&{\bf 8}
& {\bf < 6}&{\bf 6} & {\bf 7}&{\bf 6} \cr
\phantom{LAB\,1}b\ldots\ldots\ldots    & 1&5    & $-$3&4 & 5&5 & 7&3 & 7&6 & 7&9\cr
\phantom{LAB\,1}c\ldots\ldots\ldots    & 1&5    & 10&7   & 4&8 & 3&8 & $<$6&6 & $<$6&6 \cr
\phantom{LAB\,1}d\ldots\ldots\ldots   & $-$3&0 & 7&7    & 2&8 & 1&9 & $<$6&6 & $<$6&6\cr
\phantom{LAB\,1}e\ldots\ldots\ldots    & $-$4&5 & $-$1&4 & 1&5 & $<$0&9 & $<$6&6 & $<$6&6\cr
LAB\,2a\ldots\ldots\ldots             & 3&0    & $-$7&7 & 3&4 & 4&1 & $<$6&6 & $<$6&6\cr
\phantom{LAB\,2}{\bf b}\ldots\ldots\ldots    &
{\bf 0}&{\bf 0}    & {\bf 0}&{\bf 0}    & {\bf 3}&{\bf 8} & {\bf 5}&{\bf 0} &
{\bf <6}&{\bf 6} & {\bf 9}&{\bf 2}\cr
\phantom{LAB\,2}c\ldots\ldots\ldots     & $-$3&0 & $-$5&5 & 16&3 & 12&0 & 8&1 & $<$6&6\cr
\hline
\end{tabular}
\tablecomments{(1) Positional offsets from the SMGs in LAB\,1 and LAB\,2 
identified in \S3.
The proposed mid-infrared counterparts, LAB\,1a and LAB\,2b are at
$22^{\rm h}17^{\rm m}26.0^{\rm s}$, $+00^\circ12'36.2''$ and
  $22^{\rm h}17^{\rm m}39.1^{\rm s}$, $+00^\circ13'30.4''$ (J\,2000)
  respectively.
  Positional uncertainties based on the 3.6$\mu$m centroids are 0.8$''$. (2)
Flux uncertainties are 0.3$\mu$Jy in 3.6$\mu$m \& 4.5$\mu$m and 2.2$\mu$Jy in
5.8$\mu$m \& 8.0$\mu$m. Quoted limits are 3-$\sigma$.}
\end{inlinetable}

To further test if the optical-mid-infrared SEDs of the sources can be
reproduced by stellar populations, we use 
{\sc hyperz} (Bolzonella, Miralles \& Pello\ 2000) to model the stellar
continuum by fitting to the rest-frame 0.15--2\micron photometry. 
We adopt a model with an exponentially declining SFR, with $\tau=30$\,Gyr, from 
Bruzual \& Charlot (1993) and constrain the
redshift to be $z=3.09$. Dust is modelled with a Calzetti 
law (Calzetti \etal 2000). The rest-frame optical--near-infrared portions of
the SEDs are well fit by this model, with high-reddening ($A_V\sim 3$) and young ages. 
Absolute $K$-band magnitudes based on the fit are 
$M_K \sim -26.8$ in LAB\,1a and $M_K \sim -27.2$ for LAB\,2b. These 
absolute rest-frame $K$-band magnitudes are comparable to the wider SMG
population (Borys et al.\ 2005), and suggest stellar masses of 
$M_\star \sim 10^{11}$\,$M_\odot$, although the
$K$-band magnitude of LAB\,2b is likely to be contaminated by AGN emission,
with a significant rest-frame 2\micron excess over the 
{\sc hyperz} fit of $\sim$5$\mu$Jy. 
The detection of X-ray emission from  LAB\,2b is the best evidence
we have that it contains an AGN. To quantitatively test the correspondence between
the mid-infrared excess and the X-ray emission we use the rest-frame 
X-ray--10.5$\mu$m
correlation for AGN from Krabbe \etal (2002). 
We predict the rest-frame 10.5$\mu$m emission
for LAB\,2b by extrapolating the rest-frame 2$\mu$m using the measured spectral
slope of $\alpha=-1$. With this approach we estimate the X-ray luminosity from
the AGN to be of order $10^{44}$\,erg\,s$^{-1}$, in reasonable agreement with
that observed (Basu-Zych \&
Scharf 2004). The luminosity of the X-ray source is 
consistent with a moderate contribution to the bolometric emission from an AGN 
(Alexander \etal 2005b), however it is still not clear if an AGN or starburst is
responsible for powering the extended Lyman-$\alpha$ emission.  
LAB\,1a's rest-frame $K$-band excess over the best fit stellar
template, of $\sim$3$\pm$2\,$\mu$Jy is less
significant. Nevertheless, taking the same approach as for LAB\,2b, we predict an
X-ray luminosity of $\sim$$5\times10^{43}$\,erg\,s$^{-1}$,
consistent with the 2--10\,keV limit in the existing {\it Chandra} observations.

To make progress we need mid-infrared observations of a larger sample
of LABs in SA\,22,
four of which also contain SMGs (Geach \etal 2005). This will allow us to 
investigate whether they also exhibit rest-frame near-infrared characteristics 
similar to the general submillimeter population. 
Searches for obscured AGN in SA\,22 will also be significantly advanced by 
a recently approved 400\,ks {\it Chandra} ACIS-I exposure of this region.
Together these studies will provide a clearer view of the energetics of LABs
and their relationship to the wider SMG population.

We thank an anonymous referee for helpful comments which greatly improved the
clarity of this work. We also appreciate useful discussions with Richard Wilman, 
Mark Swinbank, Yuichi Matsuda and Toru Yamada. 
J.E.G. and J.P.S. thank the
U.~K. Particle Physics and Astronomy Research Council for financial
support. I.R.S. and D.M.A. acknowledge the Royal Society.

\end{document}